# An axially-variant kernel imaging model applied to ultrasound image reconstruction

Mihai I. Florea, Adrian Basarab, Denis Kouamé, and Sergiy A. Vorobyov

*Abstract*—Existing ultrasound deconvolution approaches unrealistically assume, primarily for computational reasons, that the convolution model relies on a spatially invariant kernel and circulant boundary conditions. We discard both restrictions and introduce an image formation model applicable to ultrasound imaging and deconvolution based on an axially varying kernel, that accounts for arbitrary boundary conditions. Our model has the same computational complexity as the one employing spatially invariant convolution and has negligible memory requirements. To accommodate state-of-the-art deconvolution approaches when applied to a variety of inverse problem formulations, we also provide an equally efficient adjoint expression of our model. Simulation results confirm the tractability of our model for the deconvolution of large images. Moreover, the quality of reconstruction using our model is superior to that obtained using spatially invariant convolution.

## I. INTRODUCTION

Ultrasound imaging is a medical imaging modality widely adopted due to its efficiency, low cost, and safety. These advantages come at the expense of image quality. Consequently, the accurate estimation of the tissue reflectivity function (TRF) from ultrasound images is the subject of active research. Generally, existing approaches assume that the formation of ultrasound images follows a 2D convolution model between the TRF and the system kernel. The convolution model is further constrained for computational reasons to have spatially invariant kernel and circulant boundary conditions (see e.g. [1]–[6]).

Pulse-echo emission of focused waves still remains the most widely used acquisition scheme in ultrasound imaging. It consists of sequentially transmitting narrow focused beams. For each transmission centered at a lateral position, the raw data is used to beamform one RF signal. Given the repeatability of the process in the lateral direction, the kernels do not vary laterally. However, despite dynamic focusing in reception and time gain compensation, the kernels become wider as we move away from the focal depth, thus degrading the spatial resolution and motivating the proposed kernel variation model.

Previous works account for this variation by assuming local regions of kernel invariance and performing deconvolution block-wise (e.g. [7]). Very recently, ultrasound imaging convolution models with continuously varying kernels were proposed in [8] and [9]. However, [8] makes the overly restrictive assumption that the spatially varying kernel is obtained from a constant reference kernel modulated by the exponential of a fixed discrete generator scaled by the varying kernel center image coordinates. Therefore, it does not take into account the depth-dependent spatial resolution degradation explained previously. On the other hand, the deconvolution proposed in [9] has an iteration complexity proportional to the cube of the number of pixels in the image, limiting its applicability to very small images.

The contributions of this work are as follows. (i) We propose a novel axially-variant kernel ultrasound image formation model (Section III). (ii) Our model is linear and may be implemented as a matrix. However, the matrix form does not scale because its complexity is proportional to the square of the number of pixels in the image. Therefore we provide an efficient matrix-free implementation of axially varying convolution that entails the same computational cost as spatially invariant convolution (Subsection III-B). (iii) The deconvolution problem is ill-posed and many deconvolution models can only be solved approximately using proximal splitting methods (see [10], [11], and references therein) that compute at every iteration the gradient of a data fidelity term. The data fidelity gradient expression includes calls to both the model operator and its adjoint. We express this adjoint operator in a form of equal complexity to that of the forward model operator (Section IV). (iv) We confirm using simulation results that deconvolution with our model is tractable even for large images and produces results superior to those obtained using the spatially invariant model (Section V).

## II. NOTATION

In this work, images (ultrasound images and TRFs) are vectorized in column-major order but referenced in 2D form. For instance, image $v \in \mathbb{R}^{m_v n_v}$ corresponds to a $m_v \times n_v$ 2D image and has the pixel value at coordinates $(i, j)$ given by $v_{m_v(j-1)+i}$. However, for clarity of exposition, we denote it as 2D object $\boldsymbol{v} \in \mathbb{R}^{m_v \times n_v}$, with the pixel value at location $(i, j)$ given by $\boldsymbol{v}_{i,j}$. Bold marks this artificial indexing. Similarly, linear operators are matrices but referred to as 4D tensors, e.g., $\boldsymbol{O} : \mathbb{R}^{m_v \times n_v} \to \mathbb{R}^{m_w \times n_w}$ denotes $O \in \mathbb{R}^{m_v n_v \times m_w n_w}$.

In the sequel, we define several classes of linear operators that constitute the mathematical building blocks of our proposed model and analysis. Note that these are more general than normal linear operators because their dimensions not

M. I. Florea and S. A. Vorobyov are with Aalto University, Dept. Signal Processing and Acoustics, FI-00076, AALTO, Finland (e-mail: mihai.florea@aalto.fi; sergiy.vorobyov@aalto.fi). A. Basarab and D. Kouamé are with IRIT, CNRS UMR 5505, University of Toulouse (e-mail: basarab@irit.fr; kouame@irit.fr). Part of this work has been supported by the Academy of Finland, under grant no. 299243, and CIMI Labex, Toulouse, France, under grant ANR-11-LABX-0040-CIMI within the program ANR-11-IDEX-0002-02.

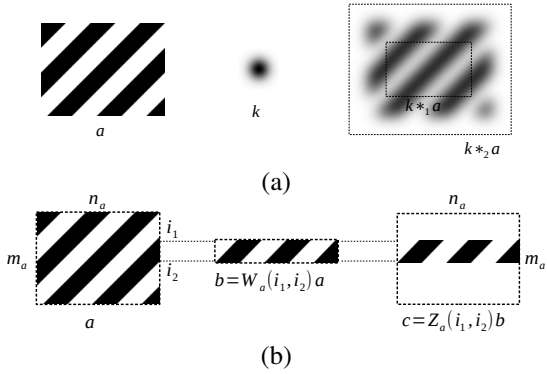

Fig. 1. (a) Convolving test image $\boldsymbol{a}$ with a Gaussian kernel $\boldsymbol{k}$. The inner rectangle represents valid convolution whereas the outer marks full convolution; (b) Applying the full-width window operator, followed by a full-width zero-padding operator on a test image $\boldsymbol{a}$. Here, black and white correspond to values of 1 and 0, respectively. Kernel $\boldsymbol{k}$ is displayed after min-max normalization.

only depend on those of their parameters but also of their arguments.

*A. Convolution operators*

For all $m_k, n_k \geq 1$, all kernels $\boldsymbol{k} \in \mathbb{R}^{m_k \times n_k}$, and all $m_a \geq m_k$, $n_a \geq n_k$, we define the linear operators $\boldsymbol{\mathcal{C}}_1(\boldsymbol{k})$ and $\boldsymbol{\mathcal{C}}_2(\boldsymbol{k})$ as

$$\boldsymbol{\mathcal{C}}_1(\boldsymbol{k})\boldsymbol{a} \stackrel{\text{def}}{=} \boldsymbol{k} *_1 \boldsymbol{a}, \quad \boldsymbol{\mathcal{C}}_2(\boldsymbol{k})\boldsymbol{a} \stackrel{\text{def}}{=} \boldsymbol{k} *_2 \boldsymbol{a},$$

for all $\boldsymbol{a} \in \mathbb{R}^{m_a \times n_a}$, where operations $*_1$ and $*_2$ denote (discrete) valid convolution and full convolution, respectively, defined as

$$(\boldsymbol{k} *_1 \boldsymbol{a})_{i,j} \stackrel{\text{def}}{=} \sum_{p=1}^{m_k} \sum_{q=1}^{n_k} \boldsymbol{k}_{p,q} \boldsymbol{a}_{i-p+m_k, j-q+n_k},$$
$$i \in \{1, ..., m_a - m_k + 1\}, \ j \in \{1, ..., n_a - n_k + 1\},$$
$$(\boldsymbol{k} *_2 \boldsymbol{a})_{i,j} \stackrel{\text{def}}{=} \sum_{p=p_i}^{\bar{p}_i} \sum_{q=q_j}^{\bar{q}_j} \boldsymbol{k}_{p,q} \boldsymbol{a}_{i-p+1, j-q+1},$$
$$i \in \{1, ..., m_a + n_k - 1\}, \ j \in \{1, ..., n_a + n_k - 1\},$$
$$p_i = \max\{1, i - m_a + 1\}, \quad \bar{p}_i = \min\{i, m_k\},$$
$$q_j = \max\{1, j - n_a + 1\}, \quad \bar{q}_j = \min\{j, n_k\}.$$

The difference between the two forms of convolution is exemplified in Fig. 1(a). Valid convolution is thereby the subset of full convolution where every output pixel is expressed using the entire kernel $\boldsymbol{k}$.

*B. Auxiliary operators*

For conciseness, we also introduce the following auxiliary operators. None involve any computation in practice.

Let the rotation operator $\boldsymbol{\mathcal{R}}(\boldsymbol{k})$ be given by

$$(\boldsymbol{\mathcal{R}}(\boldsymbol{k}))_{i,j} \stackrel{\text{def}}{=} \boldsymbol{k}_{m_k - i + 1, n_k - j + 1},$$
$$i \in \{1, ..., m_k\}, \quad j \in \{1, ..., n_k\}.$$

To further simplify notation, for all indices $1 \leq a \leq b \leq c$, we denote the exception index set $\mathcal{I}(a, b, c)$ as

$$\mathcal{I}(a, b, c) \stackrel{\text{def}}{=} \{1, ..., c\} \setminus \{a, ..., b\}.$$

The full-width window and zero padding operators are defined as

$$(\boldsymbol{\mathcal{W}}_s(i_1, i_2)\boldsymbol{a})_{i,j} \stackrel{\text{def}}{=} \boldsymbol{a}_{i+i_1, j}, \quad i \in \{0, ..., i_2 - i_1\},$$
$$(\boldsymbol{\mathcal{Z}}_s(i_1, i_2)\boldsymbol{a})_{i,j} \stackrel{\text{def}}{=} \begin{cases} \boldsymbol{a}_{i-i_1, j}, & i \in \{i_1, ..., i_2\}, \\ 0, & i \in \mathcal{I}(i_1, i_2, m_s), \end{cases}$$

where $j \in \{1, ..., n_s\}$ and index $s \in \{t, p\}$ stands for image size quantities $m_t$, $m_p$, $n_t$, and $n_p$. Their effect on a test image is shown in Fig. 1(b).

### III. AXIALLY-VARIANT KERNEL BASED ULTRASOUND IMAGING MODEL

We propose the following image formation model

$$\boldsymbol{y} = \boldsymbol{HPx} + \boldsymbol{n}, \tag{1}$$

where $\boldsymbol{x}, \boldsymbol{y}, \boldsymbol{n} \in \mathbb{R}^{m_t \times n_t}$ denote the TRF to be recovered, the observed radio-frequency (RF) image, and independent identically distributed (i.i.d.) additive white Gaussian noise (AWGN), respectively.

*A. Padding*

Operator $\boldsymbol{P} : \mathbb{R}^{m_t \times n_t} \to \mathbb{R}^{m_p \times n_p}$ pads the TRF with a boundary of width $n_r$ and height $m_r$, yielding an image of size $m_p = m_t + 2m_r$ times $n_p = n_t + 2n_r$. Padding in our ultrasound imaging model allows us to reconstruct a TRF of the same size as the observed RF image. To this end, we must simulate the effect the surrounding tissues have on the imaged tissues. Padding is an estimation of the surrounding tissues using information from the imaged TRF. This estimation only affects the border of the reconstructed TRF. If this border information is not required, the reconstructed TRF can simply be cropped accordingly. The addition of padding to our model brings the advantage of accommodating both options.

For computational reasons, $\boldsymbol{P}$ is assumed linear and separable along the dimensions of the image. Separability translates to $\boldsymbol{P} = \boldsymbol{P}_m \boldsymbol{P}_n$. Here, $\boldsymbol{P}_m$ pads every column of the image independently by applying the 1D padding (linear) operator $\boldsymbol{\mathcal{P}}(m_t, m_r)$. Consequently, when $n_t = 1$ and $n_r = 0$, operators $\boldsymbol{P}$ and $\boldsymbol{\mathcal{P}}(m_t, m_r)$ are equivalent. The row component $\boldsymbol{P}_n$ treats every row as a column vector, applies $\boldsymbol{\mathcal{P}}(n_t, n_r)$ to it, and turns the result back into a row.

Padding, either in 1D or 2D, can be performed without explicitly deriving an operator matrix. However, the matrix form facilitates the formulation of the corresponding adjoint operator. Common matrix forms of operator $\boldsymbol{\mathcal{P}}(m_t, n_t)$ are shown in Fig. 2 for $m_t = 10$ and $m_r = 3$. These examples demonstrate that the matrix form of $\boldsymbol{\mathcal{P}}(m_t, m_r)$ can be easily generated programatically and, due to its sparsity, can be stored in memory even for very large values of $m_t$ and $m_r$. These properties extend to the matrix form of the 2D padding operator $\boldsymbol{P}$ by virtue of the following result.

**Theorem 1.** *Padding operator $\boldsymbol{P}$ can be obtained programatically in the form of a sparse matrix as*

$$\boldsymbol{P} = \boldsymbol{\mathcal{P}}(n_t, n_r) \otimes \boldsymbol{\mathcal{P}}(m_t, m_r),$$

*where $\otimes$ denotes the Kronecker product.*

   *Proof:* See Appendix A. ∎

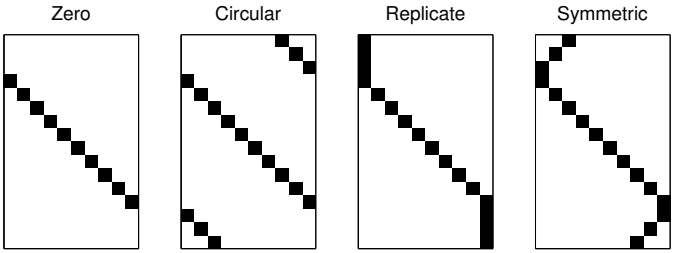

Fig. 2. Common matrix forms of 1D padding operators $\boldsymbol{\mathcal{P}}(10,3)$. Black denotes a value of 1 and white denotes 0.

### B. Axially varying convolution

Linear operator $\boldsymbol{H} : \mathbb{R}^{m_p \times n_p} \to \mathbb{R}^{m_t \times n_t}$ performs the axially-variant kernel convolution. We define axially-variant convolution as the linear operation whereby each row $i_h \in \{1, ..., m_t\}$ of the output image is obtained by the valid convolution between the kernel pertaining to that row $\boldsymbol{k}(i_h) \in \mathbb{R}^{m_k \times n_k}$, where $m_k = 2m_r + 1$ and $n_k = 2n_r + 1$, and the corresponding patch in the input (padded) TRF. The auxiliary operators defined in Section II enable us to write $\boldsymbol{H}$ as a sum of linear operators based on the observation that the concatenation of output rows has the same effect as the summation of the rows appropriately padded with zeros. Analytically, this translates to

$$\boldsymbol{H} = \sum_{i_h=1}^{m_t} \boldsymbol{\mathcal{Z}}_t(i_h, i_h) \boldsymbol{\mathcal{C}}_1(\boldsymbol{k}(i_h)) \boldsymbol{\mathcal{W}}_p(i_h, i_h + 2m_r). \quad (2)$$

In matrix form, operator $\boldsymbol{H}$ would need to store $m_p n_p m_t n_t$ coefficients and its invocation would entail an equal number of multiplications. Its complexity would thus be greater than the square of the number of pixels in the image, limiting its applicability to medium sized images. Using the matrix-free expression in (2), operator $\boldsymbol{H}$ performs $m_k n_k m_t n_t$ multiplications and has negligible memory requirements. Therefore, in ultrasound imaging, the matrix-free representation is not only vastly superior to its matrix counterpart (because the kernel is much smaller than the image), but has the same computational complexity as spatially invariant convolution (excluding the unrealistic circulant boundary case).

Unlike the forward model which, by utilizing operators $\boldsymbol{H}$ and $\boldsymbol{P}$, can be computed exactly with great efficiency, many deconvolution models can only be solved approximately using proximal splitting methods that optimize an objective containing a data fidelity term $\phi(\boldsymbol{H}\boldsymbol{P}\boldsymbol{x} - \boldsymbol{y})$. These methods employ at every iteration the gradient of the data fidelity term, given by

$$\nabla(\phi(\boldsymbol{H}\boldsymbol{P}\boldsymbol{x} - \boldsymbol{y})) = \boldsymbol{P}^T \boldsymbol{H}^T (\nabla \phi)(\boldsymbol{H}\boldsymbol{P}\boldsymbol{x} - \boldsymbol{y}). \quad (3)$$

Note that, under our AWGN assumption, $\phi$ is the square of the $\ell_2$-norm but the results in this work may be applied to other additive noise models.

The gradient expression in (3) depends on $\boldsymbol{H}$ and $\boldsymbol{P}$ as well as their adjoints. In the following, we derive computationally efficient expressions for adjoint operators $\boldsymbol{H}^T$ and $\boldsymbol{P}^T$.

### IV. ADJOINT OF MODEL OPERATOR

By taking the adjoint in (2), we get

$$\boldsymbol{H}^T = \sum_{i_h=1}^{m_t} (\boldsymbol{\mathcal{W}}_p(i_h, i_h + 2m_r))^T (\boldsymbol{\mathcal{C}}_1(\boldsymbol{k}(i_h)))^T (\boldsymbol{\mathcal{Z}}_t(i_h, i_h))^T.$$

To obtain a matrix-free representation of $\boldsymbol{H}^T$, we need the corresponding matrix-free expression for the adjoints of the convolution and auxiliary operators. First, it trivially holds that the window operator and corresponding zero padding operator are mutually adjoint, expressed as

$$(\boldsymbol{\mathcal{W}}_s(i_1, i_2))^T = \boldsymbol{\mathcal{Z}}_s(i_1, i_2). \quad (4)$$

The adjoint of valid convolution can be linked to full convolution as follows.

**Theorem 2.** *The adjoint of valid convolution is full correlation (convolution with the rotated kernel), namely*

$$(\boldsymbol{\mathcal{C}}_1(\boldsymbol{k}))^T = \boldsymbol{\mathcal{C}}_2(\boldsymbol{\mathcal{R}}(\boldsymbol{k})).$$

*Proof:* See Appendix B. ∎

Theorem 2 and (4) yield a matrix-free expression for $\boldsymbol{H}^T$ in the form of

$$\boldsymbol{H}^T = \sum_{i_h=1}^{m_t} \boldsymbol{\mathcal{Z}}_p(i_h, i_h + 2m_r) \boldsymbol{\mathcal{C}}_2(\boldsymbol{\mathcal{R}}(\boldsymbol{k}(i_h))) \boldsymbol{\mathcal{W}}_t(i_h, i_h). \quad (5)$$

Therefore, operators $\boldsymbol{H}$ and $\boldsymbol{H}^T$ have equal computational complexity. Moreover, they exhibit two levels of parallelism. The convolution operators themselves are fully parallel and the computations pertaining to each row $i_h$ can be performed concurrently. Thus, in matrix-free from, both operators benefit from parallelization in the same way as their matrix counterparts.

The adjoint of the padding operator, $\boldsymbol{P}^T$, can be obtained either directly through sparse matrix transposition or by applying transposition in Theorem 1 as

$$\boldsymbol{P}^T = (\boldsymbol{P})^T = (\boldsymbol{\mathcal{P}}(n_t, n_r))^T \otimes (\boldsymbol{\mathcal{P}}(m_t, m_r))^T. \quad (6)$$

Finally, note that whereas the column-major order assumption can be made without loss of generality for operator $\boldsymbol{H}$, it is not the case for the padding operator $\boldsymbol{P}$. In particular, row-major order reverses the terms in the Kronecker product.

### V. EXPERIMENTAL RESULTS

We have tested our model on a simulated ultrasound image deconvolution problem. Three TRFs (TRF 1, TRF 2, and TRF 3) were generated following the procedure used in computing the kidney phantom within the Field II simulator [12], [13]. Each TRF is an interpolation of Gaussian distributed random scatterers with standard deviations determined by a pixel intensity map. The map for TRF 1 is a patch from a human kidney tissue optical scan included in the Field II distribution. The maps for TRF 2 and TRF 3 are patches from a single slice (visual identifier 3272) of the female dataset provided by the Visible Human Project [14]. Deconvolution experiments were performed for each of the 3 TRFs, with the padding size, matching the kernel radii, given by $m_r \times n_r$. The TRFs are shown in Fig. 3(a1), (a2), and (a3), respectively, and their parameters are listed in Table I.

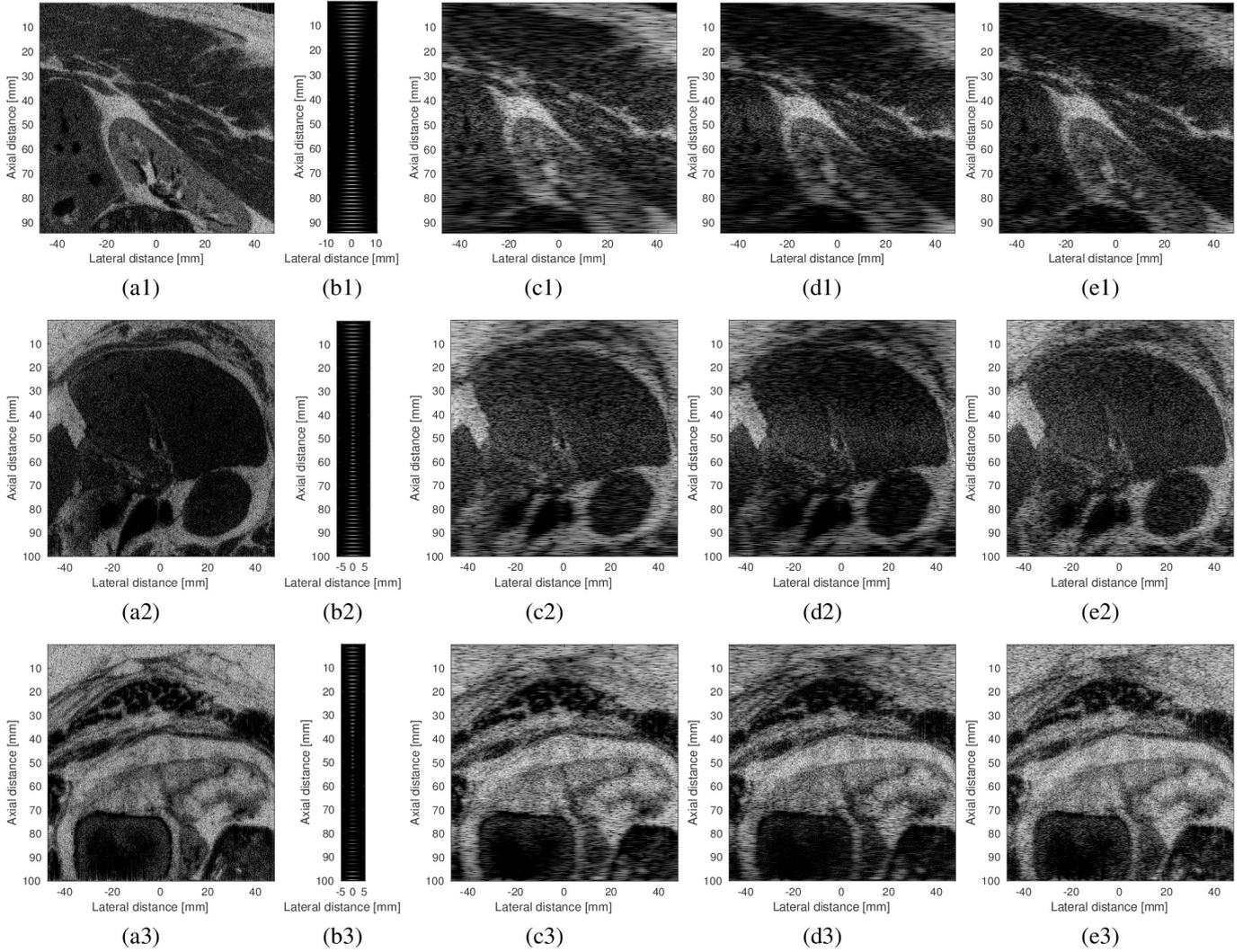

Fig. 3. (a) Ground truth (in B-mode) of the tissue reflectivity function (TRF); (b) Demodulated kernels $\boldsymbol{k}(i_h)$ for twenty depths at regularly spaced intervals of 2 mm; (c) Observed B-mode image simulated following the proposed axially-variant convolution model; (d) Axially-invariant deconvolution result AI (in B-mode) obtained with a fixed kernel equal to $\boldsymbol{k}(m_t/2)$ (the center kernel of the axially-variant model); (e) Axially-variant deconvolution result AV (in B-mode) using our model. All images are displayed using a dynamic range of 40 dB. Row (a1)-(e1) pertains to TRF 1, row (a2)-(e2) to TRF 2, and row (a3)-(e3) to TRF 3.

TABLE I
DECONVOLUTION EXPERIMENT PARAMETERS FOR EACH TRF

| TRF   | $m_t$ | $n_t$ | Axial size | Lateral size | $m_r$ | $n_r$ |
|-------|-------|-------|------------|--------------|-------|-------|
| TRF 1 | 2480  | 480   | 94 mm      | 95 mm        | 9     | 50    |
| TRF 2 | 2598  | 480   | 100 mm     | 95 mm        | 7     | 35    |
| TRF 3 | 2598  | 480   | 100 mm     | 95 mm        | 5     | 25    |

For every row $i_h \in \{1, ..., m_t\}$, we have defined the kernel $\boldsymbol{k}(i_h)$ in (2) as

$$\boldsymbol{k}(i_h)_{i,j} = \rho_{\mu_z,\sigma_z}(i)\rho_{\mu_x,\sigma_x(i_h)}(j)\cos(2\pi f_0/f_s(i-\mu_z)),$$

where $\rho_{\mu,\sigma}(x)$ is a normalized Gaussian window, given by $\rho_{\mu,\sigma}(x) = \frac{1}{\sqrt{2\pi}\sigma}\exp\left(-\frac{(x-\mu)^2}{2\sigma^2}\right)$, and parameters $\mu_z$ and $\mu_x$ are the center coordinates of the kernel. Axial standard deviation (SD) was set to $\sigma_z = \sigma_1$ and lateral SD to $\sigma_x(i_h) = \sqrt{((2i_h)/m_t - 1)^2(\sigma_2^2 - \sigma_1^2) + \sigma_1^2}$, with $\sigma_1 = m_r/3$ and $\sigma_2 = n_r/3$. Here, $f_0 = 3$ MHz and $f_s = 20$ MHz are the ultrasound central and sampling frequencies, respectively. For each TRF deconvolution experiment, the depth-dependent width variation of the kernel simulates the lateral spatial resolution degradation when moving away from the focus point, located in the center of the image (47 mm from the probe for TRF 1, 50 mm for TRF 2 and 3). The envelopes of these kernels at regular intervals across the image are shown in Fig. 3(b1), (b2), and (b3). Whereas the TRFs are of approximately the same size, the intensity of the blur differs in each experiment in order to display both the high and low frequency reconstruction capabilities of our method. We chose symmetric padding, as illustrated in Fig. 2, because it is more realistic than circular and zero padding and, by using a larger number of pixels from the TRF, more robust to noise than replicate padding. A small amount of noise was added to each TRF such that the signal to noise ratio is 40 dB. The ultrasound images produced from each TRF using our forward model in (1) are shown in Fig. 3(c1), (c2), and (c3).

To estimate the TRFs, we have considered an elastic net [15] regularized least-squares (based on the AWGN assumption) deconvolution model

$$\min_{\boldsymbol{x}} \frac{1}{2}\|\boldsymbol{HPx} - \boldsymbol{y}\|_2^2 + \lambda_1\|\boldsymbol{x}\|_1 + \frac{\lambda_2}{2}\|\boldsymbol{x}\|_2^2.$$

with manually tuned parameters $\lambda_1 = 2\mathrm{e}{-3}$ and $\lambda_2 = 1\mathrm{e}{-4}$. For deconvolution, we have employed the Accelerated Composite Gradient Method (ACGM) [16], [17] on account of its low resource usage, applicability, adaptability, and near-optimal linear convergence rate on elastic net regularized optimization problems.

As the term proximal splitting suggests, ACGM splits the objective into two functions

$$f(\boldsymbol{x}) = \frac{1}{2}\|\boldsymbol{HPx} - \boldsymbol{y}\|_2^2, \quad \Psi(\boldsymbol{x}) = \lambda_1\|\boldsymbol{x}\|_1 + \frac{\lambda_2}{2}\|\boldsymbol{x}\|_2^2.$$

Every iteration of ACGM is dominated by the computationally intensive data fidelity gradient function $\nabla f(\boldsymbol{x}) = \boldsymbol{P}^T\boldsymbol{H}^T(\boldsymbol{HPx} - \boldsymbol{y})$. All other calculations performed by ACGM are either negligible when compared to $\nabla f(\boldsymbol{x})$ or can be reduced to subexpressions of $\nabla f(\boldsymbol{x})$.

Due to the efficient matrix-free expressions of $\boldsymbol{H}$ in (2) and $\boldsymbol{H}^T$ in (5) as well as the sparse matrix implementation of $\boldsymbol{P}$ and $\boldsymbol{P}^T$ (easily precomputed using Theorem 1 and (6), respectively), deconvolution with our model entails the same computational cost as with a fixed kernel model.

The results of axially-invariant deconvolution (AI) are shown in Fig. 3(d1), (d2), and (d3) and using our axially-variant model (AV) in Fig. 3(e1), (e2), and (e3), all after 150 iterations. Our approach achieves almost perfect low frequency reconstruction for each TRF. For higher frequencies, blurring destroys visual information. Therefore, reconstruction quality at this level depends on the size of the kernels. For TRF 3, axial standard deviation values are the smallest and the blurring process has not suppressed all minor details. Here, our method manages to reconstruct even small, high contrast features.

For all TRFs, the gain in reconstruction quality is evident especially in the upper and lower extremities, as can be discerned from Fig. 3. Interestingly, even though the two models differ only slightly in the center of the image, our model performs better in that region as well.

## VI. Conclusions

In this work, we have proposed an axially varying convolution forward model for ultrasound imaging. The physics of ultrasound image formation as well as our deconvolution simulation results show the superiority of our model over the traditional fixed kernel model.

Our matrix-free formulas for the adjoints of the convolution and auxiliary operators, necessary for the implementation of deconvolution using proximal splitting techniques, also constitute a solid theoretical foundation for deconvolution methodologies using more sophisticated models, particularly those where the kernel also varies along the lateral direction. Furthermore, our theoretical results and methodology are not restricted to ultrasound imaging and may be extrapolated to other imaging modalities and applications.

## Appendix A
### Proof of Theorem 1

Since image columns are staked one on top of the other, $\boldsymbol{P}_m$ will have a block diagonal structure, with each block given by $\boldsymbol{\mathcal{P}}(m_t, m_r)$, namely

$$\begin{aligned}\boldsymbol{P}_m &= \mathrm{diag}\{\boldsymbol{\mathcal{P}}(m_t, m_r), \boldsymbol{\mathcal{P}}(m_t, m_r), ..., \boldsymbol{\mathcal{P}}(m_t, m_r)\} \\ &= \boldsymbol{I}_{n_t} \otimes \boldsymbol{\mathcal{P}}(m_t, m_r).\end{aligned} \quad (7)$$

Matrix $\boldsymbol{P}_n$ lacks this block structure. It instead has the elements of $\boldsymbol{\mathcal{P}}(n_t, n_r)$ strided horizontally by $m_t$ and replicated along the diagonal. This can be expressed succinctly as

$$\begin{aligned}\boldsymbol{P}_m &= \begin{bmatrix} \boldsymbol{\mathcal{P}}(n_t, n_r)_{1,1}\boldsymbol{I}_{m_t} & \cdots & \boldsymbol{\mathcal{P}}(n_t, n_r)_{1,n_t}\boldsymbol{I}_{m_t} \\ \vdots & \ddots & \vdots \\ \boldsymbol{\mathcal{P}}(n_t, n_r)_{m_p,1}\boldsymbol{I}_{m_t} & \cdots & \boldsymbol{\mathcal{P}}(n_t, n_r)_{m_p,n_t}\boldsymbol{I}_{m_t} \end{bmatrix} \\ &= \boldsymbol{\mathcal{P}}(n_t, n_r) \otimes \boldsymbol{I}_{m_t}. \end{aligned} \quad (8)$$

By combining the results along the dimensions in (7) and (8), we can compute $\boldsymbol{\mathcal{P}}$ in closed form

$$\begin{aligned}\boldsymbol{P} &= \boldsymbol{P}_m\boldsymbol{P}_n = (\boldsymbol{I}_{n_t} \otimes \boldsymbol{\mathcal{P}}(m_t, m_r))(\boldsymbol{\mathcal{P}}(n_t, n_r) \otimes \boldsymbol{I}_{m_t}) \\ &= \boldsymbol{\mathcal{P}}(n_t, n_r) \otimes \boldsymbol{\mathcal{P}}(m_t, m_r).\end{aligned} \quad (9)$$

## Appendix B
### The adjoint of valid convolution is full correlation

#### A. Additional notation

Within the context of this section, we define more general window and zero padding operators, respectively, as

$$\boldsymbol{\mathcal{W}}_{L,H}\boldsymbol{a} \stackrel{\text{def}}{=} \boldsymbol{a}_{i+m_L, j+n_L},$$
$$i \in \{0, ..., m_H - m_L\}, \quad j \in \{0, ..., n_H - n_L\},$$
$$\boldsymbol{\mathcal{Z}}_{L,H}\boldsymbol{a} \stackrel{\text{def}}{=}$$
$$\begin{cases} \boldsymbol{a}_{i-m_L, j-n_L}, & i \in \{m_L, ..., m_H\}, \quad j \in \{n_L, ..., n_H\}, \\ 0, & i \in \mathcal{I}(m_L, m_H, m_N), \quad j \in \mathcal{I}(n_L, n_H, n_N), \end{cases}$$

where indices $L, H \in \{1, k, a, M, N\}$ stand for quantities $m_1, m_k, m_a$, etc., with $m_1 = n_1 = 1$ and

$$m_M = m_a - m_k + 1, \qquad n_M = n_a - n_k + 1,$$
$$m_N = m_a + m_k - 1, \qquad n_N = n_a + n_k - 1.$$

Central to our proof is circular (periodic) convolution [18]. It only applies to arguments of equal size. When $\boldsymbol{k}$ and $\boldsymbol{a}$ are not the same size, they can only be circularly convolved by appropriately padding them with zeros. We consider arguments $\bar{\boldsymbol{k}}$ and $\bar{\boldsymbol{a}}$ of equal size $m$ by $n$.

For brevity, we introduce the circular sum $\oplus$ and difference $\ominus$, respectively given by

$$a \oplus_c b \stackrel{\text{def}}{=} ((a + b - 2) \bmod c) + 1,$$
$$a \ominus_c b \stackrel{\text{def}}{=} ((a - b) \bmod c) + 1,$$

for all $c \geq 1$ and $a, b \in \{1, ..., c\}$.

Circular convolution $\circledast$ and its corresponding linear operator $\mathcal{C}(\bar{k})$ are defined as

$$(\bar{k} \circledast \bar{a})_{i,j} \stackrel{\text{def}}{=} \sum_{p=1}^{m} \sum_{q=1}^{n} \bar{k}_{p,q} \bar{a}_{i \ominus_m p, j \ominus_n q},$$
$$i \in \{1, ..., m\}, \quad j \in \{1, ..., n\},$$
$$\mathcal{C}(\bar{k})\bar{a} \stackrel{\text{def}}{=} \bar{k} \circledast \bar{a}.$$

We further define the circular shift operator $\mathcal{S}(p,q)$ as

$$\mathcal{S}(p,q)\bar{a} \stackrel{\text{def}}{=} e(p,q) \circledast \bar{a},$$

where $e(p,q)$ is the standard basis vector image

$$e(p,q)_{i,j} = \begin{cases} 1, & i=p, \; j=q, \\ 0, & \text{otherwise}. \end{cases}$$

It follows that $\mathcal{S}(1,1)$ is the identity operator.

### B. Supporting results

**Lemma 1.** *Shift operators cumulate and can be taken out of convolutions as*

$$\mathcal{C}(\mathcal{S}(p_1,q_1)\bar{k})\mathcal{S}(p_2,q_2) = \mathcal{S}(p_1 \oplus_m p_2, q_1 \oplus_n q_2)\mathcal{C}(\bar{k}).$$

*Proof:* The result follows trivially from the commutativity and associativity properties of circular convolution

$$\mathcal{C}(\mathcal{S}(p_1,q_1)\bar{k})\mathcal{S}(p_2,q_2)\bar{a}$$
$$= (e(p_1,q_1) \circledast \bar{k}) \circledast (e(p_2,q_2) \circledast \bar{a})$$
$$= (e(p_1,q_1) \circledast e(p_2,q_2)) \circledast \bar{k} \circledast \bar{a}$$
$$= e(p_1 \oplus_m p_2, q_1 \oplus_n q_2) \circledast (\bar{k} \circledast \bar{a})$$
$$= \mathcal{S}(p_1 \oplus_m p_2, q_1 \oplus_n q_2)\mathcal{C}(\bar{k})\bar{a}.$$

∎

**Lemma 2.** *The adjoint of circular convolution is circular cross-correlation circularly shifted forward by one position*

$$(\mathcal{C}(\bar{k}))^T = \mathcal{S}(2,2)\mathcal{C}(\mathcal{R}(\bar{k})).$$

*Proof:* The convolution theorem [19] states that

$$\bar{k} \circledast \bar{a} = F^H((F\bar{k}) \odot (F\bar{a}))$$
$$= F^H \text{diag}(F\bar{k}) F \bar{a}, \qquad (10)$$

where $F$ is the Discrete Fourier Transform (DFT), $(.)^H$ denotes the Hermitian adjoint, $\odot$ represents Hadamard (element-wise) product and $\text{diag}(x)$ produces a diagonal matrix (linear operator) with the entries of $x$. Therefore, the circular convolution operator is diagonalized in Fourier domain as

$$\mathcal{C}(\bar{k}) = F^H diag(F\bar{k}) F. \qquad (11)$$

Taking the adjoint in (11), we obtain that

$$(\mathcal{C}(\bar{k}))^T = (\mathcal{C}(\bar{k}))^H = F^H diag((F\bar{k})^*) F$$
$$= F^H diag(F^*\bar{k}) F, \qquad (12)$$

where $(.)^*$ denotes the complex element-wise conjugate. The DFT matrix has conjugate symmetry [18], namely

$$F^*\bar{k} = F\mathcal{S}(2,2)\mathcal{R}(\bar{k}). \qquad (13)$$

Substituting (13) in (12) we have

$$(\mathcal{C}(\bar{k}))^T = F^H diag(F\mathcal{S}(2,2)\mathcal{R}(\bar{k})) F,$$

which can be expressed using convolution theorem (10) as

$$(\mathcal{C}(\bar{k}))^T = \mathcal{C}(\mathcal{S}(2,2)\mathcal{R}(\bar{k})). \qquad (14)$$

The application of Lemma 1 with $p_1 = q_1 = 2$, $p_2 = q_2 = 1$, in (14) completes the proof. ∎

### C. Proof of Theorem 2

All necessary theoretical results pertaining to circular convolution have already been derived. We can now revert to kernel $k$ and image $a$. Their dimensions are as specified in Section II.

Using the more general window and zero padding operators, we can express the valid and full convolution linear operators, respectively, as

$$\mathcal{C}_1(k) = \mathcal{W}_{k,a}\mathcal{C}(\mathcal{Z}_{1,k}k)\mathcal{Z}_{1,a}, \qquad (15)$$
$$\mathcal{C}_2(k) = \mathcal{W}_{1,a}\mathcal{C}(\mathcal{Z}_{1,k}k)\mathcal{Z}_{1,M}. \qquad (16)$$

Note that if $\mathcal{C}_1(k)$ takes as arguments images of size $m_a \times n_a$, its adjoint expression involves an operator $\mathcal{C}_2(k)$ with arguments of size $m_M \times n_M$. We apply the adjoint in (15) and obtain

$$(\mathcal{C}_1(k))^T = (\mathcal{Z}_{1,a})^T (\mathcal{C}(\mathcal{Z}_{1,k}k))^T (\mathcal{W}_{k,a})^T. \qquad (17)$$

Like their full-width counterparts, the window operator and corresponding zero padding operators introduced in this section are also mutually adjoint, namely

$$(\mathcal{W}_{L,H})^T = \mathcal{Z}_{L,H}. \qquad (18)$$

Applying (18) in (17) we obtain

$$(\mathcal{C}_1(k))^T = \mathcal{W}_{1,a}(\mathcal{C}(\mathcal{Z}_{1,k}k))^T \mathcal{Z}_{k,a}. \qquad (19)$$

Lemma 2 gives

$$(\mathcal{C}(\mathcal{Z}_{1,k}k))^T = \mathcal{S}(2,2)\mathcal{C}(\mathcal{R}(\mathcal{Z}_{1,k}k))$$
$$= \mathcal{S}(2,2)\mathcal{C}(\mathcal{Z}_{a,N}\mathcal{R}(k))$$
$$= \mathcal{S}(2,2)\mathcal{C}(\mathcal{S}(m_a,n_a)\mathcal{Z}_{a,N}\mathcal{R}(k))$$
$$= \mathcal{S}(m_a+1, n_a+1)\mathcal{C}(\mathcal{Z}_{1,k}\mathcal{R}(k)). \qquad (20)$$

Using (20) in (19) and applying Lemma 1 we obtain that

$$(\mathcal{C}_1(k))^T = \mathcal{W}_{1,a}\mathcal{S}(m_a+1, n_a+1)\mathcal{C}(\mathcal{Z}_{1,k}\mathcal{R}(k))$$
$$\mathcal{S}(m_k, n_k)\mathcal{Z}_{1,M}$$
$$= \mathcal{W}_{1,a}\mathcal{S}(1,1)\mathcal{C}(\mathcal{Z}_{1,k}\mathcal{R}(k))\mathcal{Z}_{1,M}$$
$$= \mathcal{W}_{1,a}\mathcal{C}(\mathcal{Z}_{1,k}\mathcal{R}(k))\mathcal{Z}_{1,M}. \qquad (21)$$

The desired result follows by observing that the right-hand sides of (21) and (16) are identical.

# REFERENCES

[1] J. Ng, R. Prager, N. Kingsbury, G. Treece, and A. Gee, "Wavelet restoration of medical pulse-echo ultrasound images in an EM framework," *IEEE Trans. Ultrason. Ferroelectr. Freq. Control*, vol. 54, no. 3, pp. 550–568, 2007.

[2] R. Rangarajan, C. V. Krishnamurthy, and K. Balasubramaniam, "Ultrasonic imaging using a computed point spread function," *IEEE Trans. Ultrason. Ferroelectr. Freq. Control*, vol. 55, no. 2, pp. 451–464, Feb. 2008.

[3] H.-C. Shin, R. Prager, J. Ng, H. Gomersall, N. Kingsbury, G. Treece, and A. Gee, "Sensitivity to point-spread function parameters in medical ultrasound image deconvolution," *Ultrasonics*, vol. 49, no. 3, pp. 344–357, 2009.

[4] M. Alessandrini, S. Maggio, J. Poree, L. D. Marchi, N. Speciale, E. Franceschini, O. Bernard, and O. Basset, "A restoration framework for ultrasonic tissue characterization," *IEEE Trans. Ultrason. Ferroelectr. Freq. Control*, vol. 58, no. 11, pp. 2344–2360, 2011.

[5] C. Dalitz, R. Pohle-Frohlich, and T. Michalk, "Point spread functions and deconvolution of ultrasonic images," *IEEE Trans. Ultrason. Ferroelectr. Freq. Control*, vol. 62, no. 3, pp. 531–544, Mar. 2015.

[6] N. Zhao, A. Basarab, D. Kouamé, and J.-Y. Tourneret, "Joint segmentation and deconvolution of ultrasound images using a hierarchical Bayesian model based on generalized Gaussian priors," *IEEE Trans. Image Process.*, vol. 25, no. 8, pp. 3736–3750, 2016.

[7] J. G. Nagy and D. P. O'Leary, "Restoring images degraded by spatially variant blur," *SIAM J. Sci. Comput.*, vol. 19, no. 4, pp. 1063–1082, Jul. 1998.

[8] O. V. Michailovich, "Non-stationary blind deconvolution of medical ultrasound scans," in *Proc. SPIE*, vol. 101391C, Mar. 2017.

[9] L. Roquette, M. M. J.-A. Simeoni, P. Hurley, and A. G. J. Besson, "On an analytical, spatially-varying, point-spread-function," in *2017 IEEE International Ultrasound Symposium (IUS)*, Sep. 2017, Washington D.C., USA.

[10] P. L. Combettes and J.-C. Pesquet, "Proximal splitting methods in signal processing," in *Fixed-point algorithms for inverse problems in science and engineering*. Springer, 2011, pp. 185–212.

[11] N. Parikh, S. P. Boyd *et al.*, "Proximal algorithms," *Found. Trends Optim.*, vol. 1, no. 3, pp. 127–239, 2014.

[12] J. A. Jensen, "Field: A program for simulating ultrasound systems," *Med. Biol. Eng. Comput.*, vol. 34, pp. 351–353, 1996.

[13] J. A. Jensen and N. B. Svendsen, "Calculation of pressure fields from arbitrarily shaped, apodized, and excited ultrasound transducers," *IEEE Trans. Ultrason. Ferroelectr. Freq. Control*, vol. 39, no. 2, pp. 262–267, Mar. 1992.

[14] M. J. Ackerman, "The visible human project," *Proceedings of the IEEE*, vol. 86, no. 3, pp. 504–511, 1998.

[15] H. Zou and T. Hastie, "Regularization and variable selection via the elastic net," *J. R. Stat. Soc. Ser. B. Methodol.*, vol. 67, no. 2, pp. 301–320, 2005.

[16] M. I. Florea and S. A. Vorobyov, "An accelerated composite gradient method for large-scale composite objective problems," *arXiv preprint arXiv:1612.02352*, 2016.

[17] ——, "A generalized accelerated composite gradient method: Uniting Nesterov's fast gradient method and FISTA," *arXiv preprint arXiv:1705.10266*, 2017.

[18] B. P. Lathi, *Signal Processing and Linear Systems*. Oxford University Press, New York, 1998.

[19] R. C. Gonzalez and R. E. Woods, *Digital Image Processing*, 3rd ed. Pearson Prentice Hall, Upper Saddle River, NJ, USA, 2008.